# AI-Augmented Visible Light Communication: A Framework for Noise Mitigation and Secure Data Transmission

A. A. Nutfaji
*Department of Electrical Engineering*
*University of Sharjah*
Sharjah, United Arab Emirates
Nutfaji@ieee.org

Moustafa Hassan Elmallah
*Department of Electrical Engineering*
*University of Sharjah*
Sharjah, United Arab Emirates
U20103047@sharjah.ac.ae

*Abstract*—This paper presents a proposed AI Deep Learning model that addresses common challenges encountered in Visible Light Communication (VLC) systems. In this work, we run a Python simulation that models a basic VLC system primarily affected by Additive White Gaussian Noise (AWGN). A Deep Neural Network (DNN) is then trained to equalize the noisy signal received and improve signal integrity. The system evaluates and compares the Bit Error Rate (BER) before and after equalization to demonstrate the effectiveness of the proposed model. This paper starts by introducing the concept of visible light communication, then it dives deep into some details about the process of VLC and the challenges it faces, shortly after we propose our project which helps overcome these challenges. We finally conclude with a lead for future work, highlighting the areas that are most suitable for future improvements.

## I. INTRODUCTION

Humans technology development is so fragile that if we faced a limitation in the data transmission rates, we would always take the easy way out and increase the transmission frequency. While all the communication systems in nature (e.g., the Human brain, Bees) transmit data on a couple of hundred Hertz only. With that said, a relevant new communication scheme topology is being developed to enhance the quality of future communication systems. Visual or visible light communication (VLC) is a wireless communication technology that operates on the visible light spectrum (350 nm- 750 nm) [1]. In general, any optically coupled transmitter and receiver fall under VLC systems, but in the scope of this paper, we poured the focus on ultra-fast data transmission that is imperceptible to the human eye.
VLC spectrum falls between 430 THz and 790 THz [2]. It uses photo diodes to transmit the information through the medium, and an image sensor to receive the message. Those unique characteristics are the reason why VLC has gained attention in recent years. The Radio Frequency spectrum is congested, and some bands require a license before operating. On the other hand, VLC bands are completely open and free. Furthermore, VLC systems are a great option when it comes to security, since the information transmitted through the light cannot penetrate walls. Making VLC a reliable and secure option for indoor applications. Regarding Energy efficiency, VLC systems are excellent in this matter, thanks to the light-emitting diode (LED) infrastructure.

*Table I Comparison between Radio frequency technologies and VLC*

|  | Wi-Fi | Bluetooth | ISM | VLC |
|---|---|---|---|---|
| Spectrum | 2.4/5 GHz | 2.4 GHz | 900 MHz | 400 THz |
| Ambient interference | Low | Low | Low | High |
| Security | Limited | Limited | Limited | High |
| Electromagnetic interference | Presented | Presented | Presented | Free |

VLC is being considered for utilization in the 6th generation technologies for telecommunication. Specifically replacing Wireless-Fidelity (Wi-Fi) with Light-Fidelity (Li-Fi), which is an optimized, reliable choice for indoor applications. Alternatively, VLC is being integrated into Vehicle-to-Vehicle communication (V2V), aiming to enhance the navigation systems and autonomous driving experience. Moreover, underwater communication faces a ton of challenges. VLC systems could be a suitable path for close-range communication, which requires high data rates and simple implementation.

## II. Background & Related Work

The fundamentals of VLC are straightforward. There is a light source that can illuminate at a very high frequency and transmit the data to users through the visual spectrum. The system may support single-user or multiple-user communication. In the multi-user case, the operation is a bit more advanced since it uses multiplexing to serve all users. Multiple access techniques include:

- TDMA (Time-Division Multiple Access)
- FDMA (Frequency-Division Multiple Access)

- SDMA (Space-Division Multiple Access)
- OFDMA (Orthogonal Frequency-Division Multiple Access)

Often OFDMA and Beam steering are used to avoid interference between users. This choice comes with many challenges, like energy efficiency and decoding complexity [3]. Moreover, depending on the application, VLC systems can be one-way (simplex) or two-way (Duplex). For instance, traffic lights and indoor positioning systems use simplex, while Li-Fi and V2V use duplex communication.

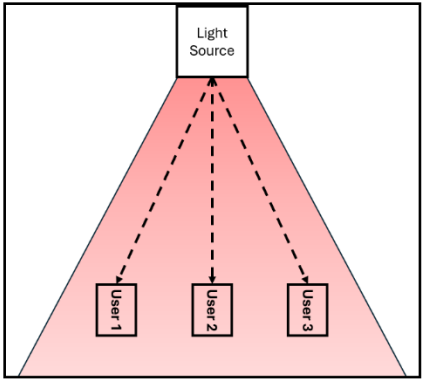

*Figure 1 Visual representation of a VLC system*

The high-level block diagram of the system is shown in Figure 2. Each block will be decomposed and analyzed to gain a deeper understanding.

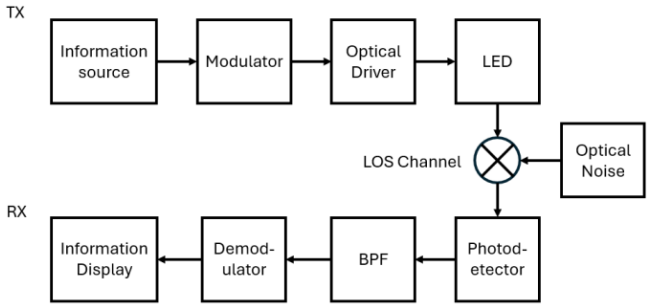

*Figure 2 VLC Block diagram, adapted from [4]*

i. Modulator

The proper modulation scheme is necessary to obtain decent results. The modulation process must be based on manipulating the aspects of light (e.g., Light intensity) to encode the message. Some popular modulation techniques may include OOK, VPPM, and CSK. For the proposed framework, OOK was chosen for the sake of simplicity. OOK is a digital modulation technique that represents the symbols by the presence or absence of the carrier signal. This technique is very suitable for the design, since the LED will have only two complementary states to operate on.

ii. Optical Driver

The optical driver is a constant current source responsible for transforming the passband signals into current and voltage signals that bias the LED. The optical driver directly influences light intensity and power efficiency.

iii. Channel

Since the photodetector area is several orders of magnitude larger than the light wavelength of the transmitted signal, the effect of multipath fading is significantly mitigated. This is due to distributing the incoming optical signal over the entire surface, which reduces the phase variations that cause the interference. However, reflected images due to objects will cause inter-symbol interference (ISI); this phenomenon requires extra attention, especially in the non-line-of-sight (NLOS) case. In the case studied, the light dispersion, in addition to the ISI effect, is modeled as a linear impulse response [5]. Hence:

$$y(t) = \gamma \, m(t) \otimes h(t) + n(t) \quad (1)$$

Where *y(t)* is the received signal, $\gamma$ is the detector responsive, *m(t)* is the transmitted message, *h(t)* is the impulse response, and $n(t) \sim \mathcal{N}(0, \sigma^2)$ Is the additive white Gaussian noise (AWGN). Overall, the noise is a mixed contribution of the following:

$$n(t) = \sigma_{shot}^2 + \sigma_{thermal}^2 + \gamma^2 P_{R,ISI}^2 \quad (2)$$

The first two terms depend on the receiver's composition and the environment. ISI noise depends on how well the pulses are shaped and how often inter-symbol interference occurs. Therefore, optimizing the pulse shaping will reduce the ISI and aid in reducing the total noise.

iv. Band-Pass Filter

Generally, filtering is done in two stages, the first stage is a physical optical filter. The second stage is an electronic filter. The purpose of filtering is to prevent saturating the photo-sensor, in addition to mitigating interference and noise.

III. PROPOSED FRAMEWORK

The scope of the framework is to apply the appropriate Artificial Intelligence (AI) method to achieve channel equalization. Specifically, diminishing BER. We propose a data-driven equalization algorithm using *Deep Learning* (DL) or, more specifically, a *hybrid neural network*. Since it combines a *Convolutional Neural Network,* which captures local patterns in the bit sequence, and then *Recurrent Neural Network,* which Learns temporal dependencies for sequence prediction and correlated bits, and the 3$^{rd}$ layer is a *Deep Neural Network,* which makes the final decision and maps the feature to the predicted probability (0 or 1). The choice was selected carefully after reviewing all AI methods. Unlike Machine learning, which requires manual function extraction, DL can train directly from the raw data. Additionally, the Scalability of DL is wider than any other alternative and can handle complex distortions and non-linearities better.

The Scope of the equalization is to estimate *m(t)* from *y(t)* stated in Eq.1 with the minimum error rate possible. Traditionally, equalizers like Zero-*Forcing* for instance, are limited by their linearity and require channel estimation preprocessing. We tackled this issue by leveraging a hybrid CNN-LSTM trained on generated pairs of *(y(t),x(t))* samples. The first layer of the model is the CNN, and it defines the 1-dimensional convolution as:

$$z_i^{(l)} = ReLU(\sum_j w_{ij}^{(l)} * z_j^{(l-1)} + b_i^{(l)}) \quad (3)$$

Where $z_i^{(l)}$ is the activation of the $i^{th}$ neuron in layer l, $w_{ij}^{(l)}$ are the trainable kernel weights, $b_i^{(l)}$ are bias terms, and Spatial locality is exploited via stride-1 convolutions with zero-padding to preserve temporal resolution. The extracted features are then passed to the second layer (LSTM units), which use gating mechanisms (Forget gate, Input gate, Candidate state, Cell state update, output gate, and Hidden state) to retain long-term dependencies:

$$\begin{aligned} f_t &= \sigma(W_f \cdot [h_{t-1}, x_t] + b_f) \\ i_t &= \sigma(W_i \cdot [h_{t-1}, x_t] + b_i) \\ \tilde{C}_t &= \tanh(W_C \cdot [h_{t-1}, x_t] + b_C) \\ C_t &= f_t * C_{t-1} + i_t * \tilde{C}_t \\ o_t &= \sigma(W_o \cdot [h_{t-1}, x_t] + b_o) \\ h_t &= o_t * \tanh(C_t) \end{aligned} \quad (4)$$

With this structure, temporal patterns are learned, and inter-symbol interference is mitigated. Finally, the Dense layer is included for classification.

## IV. IMPLEMENTATION AND EXPERIMENTAL SETUP

All experiments were conducted using Python 3 with TensorFlow and Keras backend on a machine equipped with an NVIDIA GPU. The system simulates binary communication over a visible light communication (VLC) channel and applies a neural network-based equalizer to mitigate nonlinear distortion and multipath effects.

### i. Data Generation and Channel Simulation

A total of 50,000 correlated binary bits were generated using a first-order Markov process with a transition probability of 0.2. The VLC channel includes a nonlinear LED response modeled by a sigmoid function and a simplified multipath component simulated as a delayed and scaled copy of the transmitted signal. Additive white Gaussian noise (AWGN) is added based on the specified signal-to-noise ratio (SNR) in dB. The received signal is aligned to compensate for multipath delay.

*Table II List of Parameters*

| Parameter | Value |
|---|---|
| Number of bits | 50,000 |
| Sequence length | 64 |
| SNR range | 0 dB to 20 dB (step: 2 dB) |
| Multipath delay | 2 samples |
| Multipath attenuation | 0.3 |
| LED nonlinearity | Sigmoid: 1 / (1 + exp(-5(x - 0.5))) |

### ii. Model Architecture

The neural network-based equalizer is composed of a hybrid convolutional and recurrent structure, featuring the following layers:

**Conv1D Layer 1:** 64 filters, kernel size 5, ReLU activation, L2 regularization (λ=0.01)
**Batch Normalization**
**MaxPooling1D**: Pool size 2
**Dropout**: 30%
**Conv1D Layer 2**: 128 filters, kernel size 3, ReLU activation, L2 regularization
**Batch Normalization**
**MaxPooling1D**: Pool size 2
**Dropout**: 30%
**Bidirectional LSTM**: 64 units (returning sequences)
**Bidirectional LSTM**: 32 units
**Dense Layer**: 32 units, ReLU activation
**Dropout**: 20%
**Output Layer**: Dense(1), Sigmoid activation (binary classification)

*Table III Training Configuration*

| Parameter | Value |
|---|---|
| Optimizer | Adam |
| Learning rate | 0.0001 |
| Batch size | 128 |
| Loss function | Binary Crossentropy |
| Metrics | Accuracy, Precision, Recall |
| Early stopping | Patience = 10 |
| ReduceLROnPlateau | Factor = 0.5, Patience = 3 |
| Train/Validation split | 80% / 20% |

### iii. Evaluation

The model's performance is assessed in terms of Bit Error Rate (BER) and classification accuracy over a range of SNR values. A comparison is made between hard-decision demodulation

(threshold at 0.5) and AI-based equalization. Visualization includes time-domain signal plots and BER curves pre- and post-equalization.

## V. RESULTS

The performance of the model was evaluated across different SNR levels (0-20dB) using BER. Pretraining, the BER remained static and ranged between 0.3-0.4, which indicates that the system is highly susceptible to noise and interference. Post training the CNN-LTSM model, a dramatic improvement appeared in the results. The BER exhibits a decreasing trend, reducing from around 0.35 at 0dB SNR to 0.01 at 20dB SNR. The following Figure presents the obtained results:

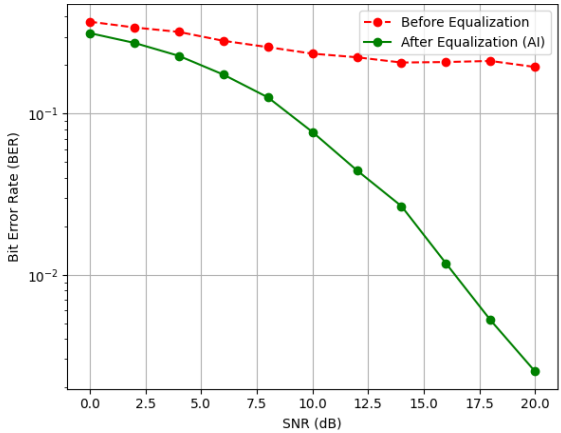

*Figure 3 BER before and after equalization*

This indicates that the model is effectively learning the inverse mapping process $f_\theta(y(t)) \to \hat{x}(t)$ with a parameter $\theta$.

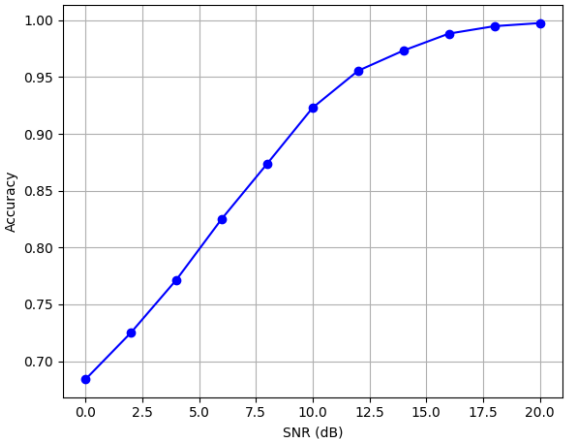

*Figure 4 Accuracy after Equalization*

Furthermore, the accuracy, defined as the percentage of correctly classified bits, presented in Figure 4, showed a solid improvement from 68% at 0dB to 99.8% at 20 dB. These findings suggest that the model is feasible for practical integration with VLC systems.

## VI. FUTURE WORK

Even while the current work demonstrates the application of deep learning, specifically a CNN-LSTM hybrid network, to counter noise and distortion of the signal in VLC systems, there are certain avenues to explore further regarding augmentation. One promising direction is extending the model to support higher-order modulation schemes beyond On-Off Keying (OOK), such as Pulse Position Modulation (PPM) or Quadrature Amplitude Modulation (QAM), which are more spectrally efficient and offer higher data rates. This would require more advanced architectures capable of supporting multi-class classification or regression outputs.

Furthermore, the simulation setup could be enhanced to incorporate a realistic physical layer that includes ambient light interference, mobility-induced fading, and device misalignment. Reinforcement learning for dynamic adaptation to time-varying channels and changing SNR levels could also be part of future work.

## ACKNOWLEDGMENT

The authors would like to specially thank the authors of the initial open-source code, which was hosted in the Project_1_AI_Receiver_CNN GitHub repository. The original code was created to mitigate noise and evaluate distorted signals, and estimate bit recovery accuracy. But it was repurposed to fit the presented application.

We also want to express our sincere thanks to Dr. Khawla Alnajjar for her invaluable guidance, which illuminated our research journey. The collaborative efforts of our dedicated team members and advisors have profoundly influenced our project, highlighting the strength of teamwork. The resources and nurturing environment provided by our university played a pivotal role in enriching our research experience.